\documentclass[10pt,conference]{IEEEtran}
\IEEEoverridecommandlockouts

\usepackage{cite}
\usepackage{amsmath,amssymb,amsfonts}
\usepackage{algorithmic}
\usepackage{graphicx}
\usepackage{textcomp}
\usepackage{xcolor}
\usepackage{listings}
\usepackage{tabularx}

\usepackage{multirow}
\usepackage{syntax}
\usepackage{url}

\usepackage{subfigure}

\usepackage{boxedminipage2e}

\colorlet{punct}{red!60!black}
\definecolor{background}{HTML}{EEEEEE}
\definecolor{delim}{RGB}{20,105,176}
\colorlet{numb}{magenta!60!black}

\lstdefinelanguage{fd2IoT}{
    basicstyle=\ttfamily\scriptsize,
    numbers=left,
    numberstyle=\scriptsize,
    stepnumber=1,
    numbersep=8pt,
    xleftmargin=3.5ex,
    showstringspaces=false,
    breaklines=true,
    frame=lines,
    backgroundcolor=\color{background},
    literate=
     *{scenario}{{{\color{numb}scenario}}}{8}
      {saturate}{{{\color{numb}saturate}}}{8}
      {alter}{{{\color{numb}alter}}}{5}
      {delete}{{{\color{numb}delete}}}{6}
      {where}{{{\color{numb}where}}}{5}
      {shake}{{{\color{numb}shake}}}{5}
      {geolocation}{{{\color{numb}geolocation}}}{11}
      {ticker}{{{\color{numb}ticker}}}{6}
      {set}{{{\color{numb}set}}}{5}
      {from}{{{\color{numb}from}}}{4}
      {with}{{{\color{numb}with}}}{4}
      {to}{{{\color{numb}to}}}{2}
      {:}{{{\color{punct}{:}}}}{1}
      {,}{{{\color{punct}{,}}}}{1}
      {;}{{{\color{delim}{;}}}}{1}
      {"}{{{\color{delim}{"}}}}{1}
      {\{}{{{\color{delim}{\{}}}}{1}
      {\}}{{{\color{delim}{\}}}}}{1}
      {[}{{{\color{delim}{[}}}}{1}
      {]}{{{\color{delim}{]}}}}{1},
}

\lstdefinelanguage{json}{
    basicstyle=\ttfamily\scriptsize,
    numbers=left,
    numberstyle=\scriptsize,
    stepnumber=1,
    numbersep=8pt,
    xleftmargin=3.5ex,
    showstringspaces=false,
    breaklines=true,
    frame=lines,
    backgroundcolor=\color{background},
    literate=
     *{0}{{{\color{numb}0}}}{1}
      {1}{{{\color{numb}1}}}{1}
      {2}{{{\color{numb}2}}}{1}
      {3}{{{\color{numb}3}}}{1}
      {4}{{{\color{numb}4}}}{1}
      {5}{{{\color{numb}5}}}{1}
      {6}{{{\color{numb}6}}}{1}
      {7}{{{\color{numb}7}}}{1}
      {8}{{{\color{numb}8}}}{1}
      {9}{{{\color{numb}9}}}{1}
      {:}{{{\color{punct}{:}}}}{1}
      {,}{{{\color{punct}{,}}}}{1}
      {\{}{{{\color{delim}{\{}}}}{1}
      {\}}{{{\color{delim}{\}}}}}{1}
      {[}{{{\color{delim}{[}}}}{1}
      {]}{{{\color{delim}{]}}}}{1},
}

\def\BibTeX{{\rm B\kern-.05em{\sc i\kern-.025em b}\kern-.08em
    T\kern-.1667em\lower.7ex\hbox{E}\kern-.125emX}}
\begin{document}

\title{A Language for Modelling False Data Injection Attacks in Internet of Things
\thanks{Work supported by the project ANR Gelead (ANR-18-ASTR-0011) and EIPHI Graduate School (contract ANR-17-
EURE-0002)}
}

\author{\IEEEauthorblockN{1\textsuperscript{st} Mathieu Briland}
\IEEEauthorblockA{\textit{DISC/FEMTO-ST} \\
\textit{University of Bourgogne Franche-Comté}\\
Besançon, France \\
mathieu.briland@univ-fcomte.fr}
\and
\IEEEauthorblockN{2\textsuperscript{nd} Fabrice Bouquet}
\IEEEauthorblockA{\textit{DISC/FEMTO-ST} \\
\textit{University of Bourgogne Franche-Comté}\\
Besançon, France \\
fabrice.bouquet@univ-fcomte.fr}
}

\maketitle

\begin{abstract}
Internet of Things (IoT) is now omnipresent in all aspects of life and provides a large number of potentially critical services. For this, Internet of Things relies on the data collected by objects. Data integrity is therefore essential. Unfortunately, this integrity is threatened by a type of attack known as False Data Injection Attack. This consists of an attacker who injects fabricated data into a system to modify its behaviour. In this work, we dissect and present a method that uses a Domain-Specific Language (DSL) to generate altered data, allowing these attacks to be simulated and tested.

\end{abstract}


\section{Introduction}

This work is carried out within Flowbird company\footnote{\url{http://www.flowbird.group}}. Flowbird is the world leader in on-street parking solutions. Parking meters manufactured by the company are present in hundreds of cities around the world and process hundreds of thousands of data in order to provide user services such as parking or environmental monitoring.

Parking meters are internet-connected devices and share the same characteristics as Internet of Things (IoT) devices such as their architecture, physical vulnerability or services provided. They can be considered as part of the IoT devices. They are therefore susceptible to being attacked like any IoT devices.

\subsection{Internet of Things}

The constant increase in the adoption of IoT makes it omnipresent in all spheres of our lives. Whether in private or public environments, whether in our homes for domestic services or in hospitals for vital services, we are surrounded by IoT devices. This growing adoption brings us many challenges in terms of security and privacy, particularly because of the very nature of IoT.

By nature IoT is a network of networks, it is composed of many heterogeneous devices (things) and technologies. The main idea behind the IoT is to provide an interface between the virtual world and the physical world, allowing the recovery, the transfer, the storage and the processing of the data gathered by the things. The data collected within the IoT are therefore critical to its functioning. Services offered by the IoT are entirely dependent on the data collected.

Also, from an architectural point of view, IoT is often described as a five-layered architecture \cite{khan2012}. First, the perception layer which is essentially about the data collection by the devices (temperature, pollution, acceleration, etc.). The collected data are then sent to the network layer. It is connected to internet and can use different mediums, technologies and protocols to transport the data generated by the things to next layers. The middleware layer then processes the data and makes decisions on actions that need to be taken. The application layer is the front end of the data processing. It is where the data are presented through graphical representation to the end user. Finally, the business layer is used by management systems to control the IoT chain and define new business models.

Therefore, from the architecture, data are at the centre of the IoT. A problem of integrity in one data is reflected in all layers and therefore in the services provided.

A particular type of attack that is little studied in the field of IoT and which specifically targets data integrity is False Data Injection Attack (FDIA).

\subsection{False Data Injection Attack}

First introduced by \cite{Sencun2004} in the field of sensor networks and later in the field of Wireless Sensor Networks (WSN), especially smart grids \cite{liu2009}, FDIA are attacks in which an attacker seeks to change a system's behaviour by modifying the data used for its services. For example, in the case of smart grids, the attacker seeks to inject errors into sensor state variables, which then leads systems to wrong power grid state estimation. In this particular area, this can lead to total power blackouts such as the one in Ukraine in 2015, where a FDIA was launched against the country's power grid. Three energy distribution companies were compromised and their services were disrupted, throwing part of Ukraine into a blackout \cite{liang2017}. 

More generally, FDIA can be defined as a cyber-attack where an attacker, thanks to his in-depth knowledge of the system under attack, compromises it. Especially in data production devices and data manipulation software. The objective is to inject falsified and altered data, with the aim of modifying the normal behaviour of these systems. It is a discreet attack that performs small injections to avoid the various protection systems. Therefore, the injection can extend over a long period of time and is difficult to detect.

\section{Background and Related Work}
Since the first appearance of the term, FDIA research has mainly focused on the impact, filtering and detection of these attacks on multiple domains with a strong focus on smart grid. In Google scholar, when searching for publications with the query "false data injection attack(s) OR stealthy injection attack(s) OR bad data injection attack(s) OR injection false data" in their titles, 490 results are returned.

By modifying this query to look for the popularity of FDIA in different domains, we can see in Table~\ref{tab:queryscholardomain} that IoT is a very little-explored domain today. The terms used for queries are, of course, not exhaustive. Publications may escape these queries because they use synonyms or do not explicitly refer to the domain in the title. Nevertheless, we have browsed through the 490 articles presented in the first search. We can therefore say that the Table \ref{tab:queryscholardomain} has a good representation of the publications according to the domain. Other terms can for example cover the field of smart grids, such as SCADA or state estimation. However, these keywords can also be found in other domains such as WSN, which is why we do not use them in our queries. Other domains also appear in the results of the first search and are poorly explored, thus they do not appear in our table, such as multi-agent systems or health care.

\setlength{\extrarowheight}{1pt}
\begin{table}[!ht]
\centering
\caption{Query of different domain associates to FDIA}
\label{tab:queryscholardomain}
\begin{tabularx}{0.5\textwidth}{|c|X|c|}
\hline
\textbf{Domain} & \textbf{Query terms} & \textbf{result number}\\ \hline
  Smart grid & "smart-grids" OR power OR "smart-grid" OR electricity OR AC OR DC & 254 \\ \hline
  CPS / CPNS & CPS OR CPNS OR "cyber physical" OR "control systems" OR "control system" &  81 \\ \hline
  WSN & WSN OR "wireless sensor network" OR "wireless sensor networks" & 49 \\ \hline
  IoT & "Internet of things" OR "Internet of thing" OR IoT & 5 \\ \hline
    ATC & ADS-B OR ATC OR "air traffic" & 4 \\ \hline
\end{tabularx}
\end{table}
\setlength{\extrarowheight}{0pt}

In \cite{bostami2019}, authors review FDIA in IoT domain and came to the same conclusion. FDIA is a big challenge within IoT, but apart for smart grids, FDIA has been very little studied.

Overall, FDIA is a fairly popular and well-studied area in smart grids and cyberphysical system (CPS). Nevertheless, its studies focus on a few very specific areas. By deepening the exact research topics, we can see that a majority of the work focuses on detection and filtering of these attacks. 

One of the difficulties of FDIA research is to be able to develop, train and verify filtering and detection techniques using real data from systems in production that have been attacked. Usually, attacked data are either protected for confidential purposes or simply not detected, therefore not flagged as compromised. To develop their attack mitigation systems and validate them through experimentation, various authors used several methods to generate data. \cite{yang2017} uses pseudorandom generator to emulate the data collection and also a pseudorandom generator to emulate FDIA behaviour. \cite{yhuang2011} uses for their system in normal state (no attacked) a Bayesian model of the random state variables with a Gaussian distribution and for the malicious data they changed the distribution. The data used by \cite{chaojun2015} are based on the data from the New York independent system operator (NYISO) from 2012 and generate the state data following a procedure. The attacked data are numerical and they apply a modification of 90\%, 95\%, 100\%, 105\%, and 110\% of the original numerical value. The main flaw in the use of these methods is usually the loss of correlation with reality. The use of nonreal base data and arbitrarily designed attacks result in the loss of both system-specific and attacker-specific behaviour. The closest work but not in the IoT domain is made by \cite{cretin2018}, who developed a DSL-based testing framework to perform FDIA on air traffic control (ATC) systems. They use real data from air traffic control and perform FDIA on them by using a DSL adapted to the specificity of the aircraft domain. The scope of the domain is also a big difference in the way FDIA are handled, for example \cite{liu2009} relies on the number of meters and the number of state variables to find an attack vector. In IoT, FDIA will not necessarily try to disrupt the state estimation but rather to disrupt the data aggregation or decrease the quality and confidence in the data to trigger actions and events. As far as we know, there is no related work to assess the resilience of IoT systems attacked by FDIA.

\section{Approach}

\begin{figure}[!ht]
\centering
\includegraphics[trim={1.8cm 1cm 1cm 1cm},clip=true,width=0.5\textwidth]{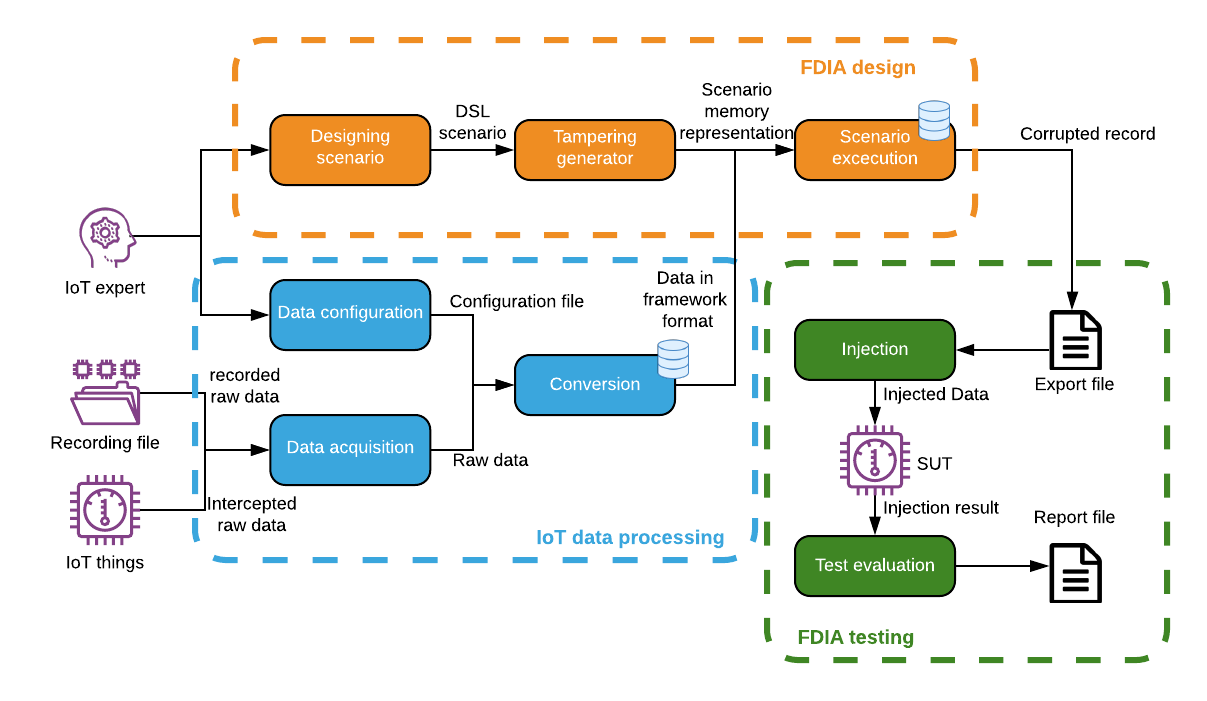}
\caption{Workflow for FDIA testing - Data acquisition, designing and testing}
\label{fig:fdiaworkflow}
\end{figure}

The aim of this work is to provide an efficient testing tool to assess the resilience of the Internet of things systems against FDIA. One of the major difficulties of this study is the total heterogeneity of IoT ecosystems. Few standards bodies have attempted to provide some standard framework for IoT (oneM2M, NGSI-LD) but in reality the vast majority of IoT follow their own protocol and data representation. Beyond their data representation, IoT systems have the same data flow through their layers, so FDIA can be performed on all types of IoT ecosystems.

\subsection{Workflow}

The workflow of our approach is shown in Fig.~\ref{fig:fdiaworkflow}.

\textbf{Data acquisition} is the first step of this workflow, it consists of two things: either importing an IoT dataset under a file format such as a CSV or JSON, or intercept the IoT dataflow by performing a man in the middle or eavesdropping attack to perform a live attack.

\textbf{Data configuration} is made by the expert of the System Under Test (SUT). It consists in the property definition of the data present in the dataset or in the flow defined in the data acquisition step.

\textbf{The conversion} is the step where input data are transformed into an internal format for better data processing and to handle a large heterogeneity of data and IoT. These converted data are stored in a database.

\textbf{Designing scenario} is made by the expert of the SUT. We provide a textual DSL for designing a scenario of FDIA. The expert uses this DSL and selects the specific dataset he wants to alter. 

\textbf{Tampering generator} is the phase where we read the scenario describing FDIA, and transform it into a memory representation.

\textbf{Scenario execution} is the process of applying the scenario written by the expert to the dataset. It results in a new dataset with tampered data stored in a database. After this step the expert can choose to export this dataset. Either for analysis, to inject into the SUT or to teach AI learning machine to detect false data injected into the dataset.

\textbf{Injection} is the process of converting the data back to their original form and injecting them into the SUT. 

\textbf{The SUT} represent the system where data altered by FDIA are injected.

\textbf{Test evaluation} aims to verify the impact of the injection in the SUT with a test oracle. This test evaluation should be exported to a report for the purpose of either correcting the SUT or improving the FDIA written by the test expert.

\section{FDIA framework}
To address this FDIA issue, we have considered the development and use of a domain-specific language (DSL). 
In \cite{deursen2020}, authors define DSL as ”a programming language or executable specification language that offers,  through appropriate notations and abstractions, expressive power focused on, and usually restricted to, a particular problem domain.”

Certainly, DSL have several drawbacks as it is a new language, training end users can be difficult. Also, building a DSL from scratch is very expensive. Whether in terms of implementation, support, maintenance or user training.
Making the choice to develop a DSL is therefore a decision that must be analysed. The problem that the DSL is supposed to solve must be analysed to see if it is an appropriate solution. 
For this purpose, we decide to follow the methodology provided in \cite{mernik2005}.
Authors define four steps: decision, analysis, design and implementation. The decision step is the answer to "When to develop a DSL" and the other steps are the answers to "How to develop a DSL".
We mainly explore the DSL through the section analysis design and implementation of the methodology.

\subsection{Domain Analysis}
\label{analysis}

In \cite{mernik2005}, authors define the analysis part as being part in which one seeks to clearly identify the field in terms of its problems and its level of knowledge. Several \textbf{input} sources can be used for this. For example, technical documentation, research carried out by experts in the field or existing code.
Also the \textbf{output} can be varied. Generally, it is a domain-specific terminology with its semantics, both formal and informal analysis are used.

Our domain, which is IoT, is technically a high level and very heterogeneous domain, taking place in a highly industralised universe. It would therefore be complicated to formally define the entire domain encompassing it. In the following, we therefore carry out an informal analysis of our domain.

\subsubsection{Scenario}

The central figure of our domain is the notion of scenarios. A security scenario is a complete step-by-step description of the actions taken by the attacker to carry out an attack. In our case we want to express the steps to simulate an FDIA attack. For this, two main elements are required. The global information of the scenario that we call the \textbf{scenario properties} and the step-by-step actions of the scenario that we call the \textbf{scenario actions}. 
A scenario is also linked to a specific record, to which it must be applied.

\subsubsection{Scenario properties}

Scenario properties are the elements that provide information about the scenario to be used globally. These properties must be scalable. According to the needs demanded by the scenario actions, new properties not thought of in the design should be easily added. This also brings the notion of mandatory property, some properties such as the scenario naming must be mandatory, while other properties can be entirely conditioned by the scenario actions, such as the use of a geolocation property.

\subsubsection{Scenario Actions}

Scenario actions are the list of actions performed by the scenario to accomplish FDIA. Scenario actions are an ordered list of actions, when interpreting them they should be read and interpreted in the order in which they appear. This implies an incidence of previous actions.
We propose to define an action by its attribute.
We identify four basic attributes for action to realise alteration attacks, called alteration primitives. They allow to define the general action that will be performed by the scenario action. Also, they define the elements that make up the action.
The four alterations primitive proposed are:
\begin{itemize}
\item \textbf{Create} is used to generate data from scratch using the information provided in the action. The generated data is inserted into the selected record. 
\item \textbf{Alter} is the action used to modify data in a record. It requires a selection criterion to select the data to be modified and an alteration criterion for modifications.
\item \textbf{Copy} is the action used to copy data from a record selected by a selection criterion. The copied data are then modified using an alteration criterion before being inserted into the original record.
\item \textbf{Delete} is the action used to delete data from a record selected by a selection criterion.
\end{itemize}

These four basic primitives can be extended later to add more complex and complete primitives requiring specific parameters to perform specific action.

After the primitives, the action is composed of multiple parts defined by the alteration primitives.

\subsubsection{Selection Criteria}

The selection criteria are used for the selection of the specific records that the attack designer wants to alter. It is the definition of the scenario action's target. Usually the selection is made through the identifier of a device, but it can also be a specific condition such as a specific value above a threshold. So it can also be seen as a trigger.

Also, specific selections need to be made depending on the type of IoT system present in the record. A specific selection can be for example, the selection inside a circle defined by geographical coordinates for its centre and a distance in metres for its radius.

\subsubsection{Time frame}

The time frame is the notion used to define the temporality of the associated action. It can be represented in two different ways. Either absolute, where we consider the first message of the recording as our timeline origin point, or in a relative way, by directly using the notion of time present in the recordings.
This property can also be seen as a specific selection criterion, it could have its place in the part of the selection criteria, nevertheless the notion of time frame is very important in the realisation of FDIA. It is therefore important that they appear in a mandatory way in the definition of an action.

\subsubsection{Alteration Criteria}

During the time frame, we define the alteration criteria. It is the effect of the action. These effects are applied to the specific records previously selected by the selection criteria. The most fundamental criterion for alteration is a simple assignment of value. The value can be a simple integer or a complex function. Nevertheless, all alteration criteria derive from an improved assignment. For example, value assignments in increments over time. So, specific alteration criteria needs to be made depending on the type of IoT system present in the record.

In the next section, we explore the third step of the methodology, the design of the DSL.

\subsection{Design}
\label{sec:design}
The design phase is the one that allows the technical definition of the language. This phase is used to define the semantics and syntax of the language taking into account the information that has been collected during the domain analysis. 

In the domain analysis section, the domain has been split into several parts. These parts can be seen in the grammar in regard of the nonterminal lexical elements. So in the next subsections we explore the design of the DSL through the different parts defined on the domain analysis. The different grammars presented in the following are in the Extended Backus-Naur Form (EBNF).

\subsubsection{Scenario}

The scenario, as stated previously, is the main part of our DSL. It represents the start symbol of the grammar. It aims to describe the complete action of a FDIA.
A scenario is a set of properties and combination of specific actions to create a FDIA. It can therefore be divided into two parts, the scenario property and the scenario actions. 
In the following, we propose the EBNF grammar of a scenario:

\begin{footnotesize}
\begin{center}
  \begin{boxedminipage}{0.9\linewidth}
\begin{grammar}

<scenario> ::= <scenarioDeclaration> <executionList>

<scenarioDeclaration> ::= Scenario <stringLiteral> `;' (<scenarioProperties> `;')$^*$

<scenarioProperties> ::= ticker <decimalLiteral>
\alt geolocation `(' <realLiteral> `,' <realLiteral> `)'

<executionList> ::= <execution> `;' (<execution> `;')$^*$

\end{grammar}
\end{boxedminipage}
\end{center}
\end{footnotesize}

The EBNF grammar 
is represented by the two nonterminal \texttt{$<$scenarioDeclaration$>$} and \texttt{$<$executionList$>$} derived from the production rule \texttt{$<$scenario$>$}.
The rule \texttt{$<$scenarioDeclaration$>$} consists in the information applied to the entire scenario and useful for its execution. The name of the scenario is the only mandatory property. At this moment we have defined two optional properties:
\begin{enumerate}
\item \textit{ticker} which is information for the elapsed time between two messages send by the SUT, also called time rate. Usually IoT systems send their data at a steady time rate. So we need to implement the ticker property to avoid detection when we create a new record in injection.
\item \textit{geolocation} which is a property to locate the point of application of the scenario. In our context, we only use localisation associates to the information of latitude and longitude.
\end{enumerate}
These properties can be extended by adding alternative production rules to the nonterminal \texttt{$<$scenarioProperties$>$}. For example, to add a property allowing geolocation in three dimensions, we must add a \texttt{$<$realLiteral$>$} to consider the altitude. We assume that in our context, we don't have the necessity to use any other representation than a 2D representation.

An example of a start of a scenario without the list of actions to perform would be: 
\begin{lstlisting}[language=fd2IoT,firstnumber=1]
scenario "exampleScenario" ;
ticker 2 ;
geolocation (47.237829,6.0240539) ;
\end{lstlisting}

After the scenario properties, we present the rule \texttt{$<$executionList$>$} that consists in the list applied by the scenario to perform action of FDIA.

\subsubsection{Scenario Actions}

Scenario actions are the core of the attack execution. They describe step by step the actions that the scenario must perform as following:

\begin{footnotesize}
\begin{center}
  \begin{boxedminipage}{0.9\linewidth}
  \begin{grammar}

<execution> ::= <create>
\alt <alter>
\alt <delete>
\alt <copy>

<create> ::= (create things <alterationCriteria> <timeframe>)

<alter> ::= (alter things <selectionCriteria> <alterationCriteria> <timeframe>)

<delete> ::= (delete things <selectionCriteria> <timeframe>)

<copy> ::= (copy things <selectionCriteria> <alterationCriteria> <timeframe>)
 
\end{grammar}
\end{boxedminipage}
\end{center}
\end{footnotesize}

An action is composed of several parts that can be seen on the scenario actions grammar. The first is the nature of the action, we call it the action \textbf{primitive}. The following parts of the action are defined by the primitive. As indicated in the domain analysis, we have three recurring parameters to perform FDIA. The part where we select the records to be impacted by the related action, the rule \texttt{$<$selectionCriteria$>$}. The part where we indicate alterations to be made on the previously selected records, the rule \texttt{$<$alterationCriteria$>$}. And the last part where we indicate the time window on which the action is to be applied, the rule \texttt{$<$timeframe$>$}. This is the skeleton of a scenario action:
\begin{center}
\begin{footnotesize}
\boxed{\underbrace{\text{\textcolor{purple}{alter}}}_\textrm{Primitive} \text{things} \ \underbrace{\text{\textcolor{purple}{where}} \ \text{ident} = "10"}_\textrm{Selection criteria} \underbrace{\text{\textcolor{purple}{set}} \ \text{tempTC} = 0}_\textrm{Alteration criteria}  \underbrace{\text{\textcolor{purple}{from}} \ 0 \ \text{\textcolor{purple}{to}} \ 4500}_\textrm{Time frame};}
\end{footnotesize}
\end{center}

\subsubsection{Selection Criteria}
The selection criteria allows you to select specific messages present in the record according to a query as following:

\begin{footnotesize}
\begin{center}
  \begin{boxedminipage}{0.9\linewidth}
  \begin{grammar}

<selectionCriteria> ::= where <selectionCriterion> (and <selectionCriterion>)$^*$

<selectionCriterion> ::= <id> `=' <type>
 \alt <id> `>' <type>
 \alt <id> `<' <type>
 \alt <id> `!=' <type>
 \alt <id> isInsideCircle `(' <realLiteral> `,' <realLiteral> `,' <decimalLiteral> `)'
 \alt <userFunction>
 
 \end{grammar}
 \end{boxedminipage}
 \end{center}
\end{footnotesize}

According to the rule \texttt{$<$selectionCriteria$>$}, the criteria always begin by the terminal symbol \texttt{"where"} and are composed of at least one criterion, if several criteria are present they are separated by the terminal symbol \texttt{"and"}. This symbol is self-explanatory, it represents the logical conjunction between the different selection criteria. The logical disjunction between criteria with the \texttt{"or"} operator is not implemented in the grammar. To get around this, it is necessary to add an action performing the same alteration using the distributivity of the operators. However, since actions are performed step by step in the order in which they appear, problems of consistency in the alteration may arise. It is therefore a future work for the implementation of the disjunction. If it appeared to us late, it is because in practice it is rarely needed to use a disjunction when selecting messages to be altered.

We defined four basic selection criteria and a complex one. They are the equality, the superiority, the inferiority and the difference. The complex one is a selection criterion that selects the messages within a circle defined by longitude, latitude and a radius in metres. These basic selections must be able to be enriched according to the SUT and the needs of attacks. The nonterminal \texttt{$<$userFunction$>$} represents this possibility. It is not really present in the grammar, it is present here to signify the importance of predicting the evolution of selection types.

Two different examples of selection criteria untied from a scenario action would be:
\begin{lstlisting}[language=fd2IoT,firstnumber=1]
where identifier= 42 and temperature > 451;
where location isInsideCircle(47.237829,6.0240539,500.0)
\end{lstlisting}

After the selection criteria, we present the alteration applied to the element selected.

\subsubsection{Alteration Criteria}

The alteration criteria are the parts where the scenario designer specifies the changes that will have to be made during the linked action. The described alterations will be applied to the selected messages during the selection criteria phase as following:

\begin{footnotesize}
\begin{center}
  \begin{boxedminipage}{0.9\linewidth}
\begin{grammar}

<alterationCriteria> ::= set <alterationCriterion> (and <alterationCriterion>)$^*$

<alterationCriterion> ::= <id> `=' <type>
\alt <id> `=' <evol>
\alt <id> `+=' <realLiteral> 
\alt <id> `*=' <realLiteral>
\alt <id> `+=' <evol> <attenuationCriteria>?
\alt <userFunction>

<evol> ::= `(' <decimalLiteral> `->' <decimalLiteral> `,' <decimalLiteral> `)'

<attenuationCriteria> ::= with attenuation of <realLiteral>
\end{grammar}
  \end{boxedminipage}
\end{center}
\end{footnotesize}

According to the rule \texttt{$<$alterationCriteria$>$}, criteria always start with the terminal symbol \texttt{"set"} and are composed of at least one criterion, if several criteria are present they are separated by the terminal symbol \texttt{"and"}. At that time we defined several types of alteration, such as simple affectation, increment, or multiplication increment. These alterations are always made up of three parts. The first is the identifier of the property to be altered. The second is a terminal symbol known as the operator to identify the type of alteration. The third is the effect of the alteration. An optional fourth part can exist, it is used to complete the alteration for certain attacks. For example, in our grammar, the rule \texttt{$<$attenuationCriteria$>$}, allows, according to the distance of the altered object from the scenario application point, to reduce the attack power.

This is the skeleton of an alteration:
\begin{center}
\begin{footnotesize}
\boxed{\underbrace{\text{Sound}}_\textrm{identifier} \ \underbrace{+=}_\textrm{Operator} \ \underbrace{3}_\textrm{Effect}\ \underbrace{\text{\textcolor{purple}{with attenuation of}} \ 10.0}_\textrm{Optional part}}
\end{footnotesize}
\end{center}

Obviously, as in the previous sections, the scalability of the language is important. Specific alteration criteria can be added to the grammar according to the SUT and the attacks needs. This is represented by the rule \texttt{$<$userFunction$>$}.

An example of an alteration criteria untied from a scenario action would be: 
\begin{lstlisting}[language=fd2IoT,firstnumber=1]
set temperature=42 and humidity +=(0.0->451.0,10.0) with attenuation of 10.0
\end{lstlisting}

After the two criteria, we present time frame selection.

\subsubsection{Time Frame}
The time frame is the part of an action that allows you to locate the action in time as the following production rule:

\begin{footnotesize}
\begin{center}
  \begin{boxedminipage}{0.9\linewidth}
  \begin{grammar}
<timeframe> ::= from <decimalLiteral> to <decimalLiteral>

 \end{grammar}
 \end{boxedminipage}
 \end{center}
\end{footnotesize}

According to the rule \texttt{$<$timeframe$>$}, the time frame always begins with the terminal symbol \texttt{"from"} followed by the start time and ends with the terminal symbol \texttt{"to"} followed by the end time.

\subsubsection{Grammar Utility}
In order not to overload the other sections with superfluous elements, not all the grammar rules have appeared previously. This section is therefore dedicated to grammar elements that are useful to our language but which did not find their place before. It is in particular about data types as following:

\begin{footnotesize}
\begin{center}
  \begin{boxedminipage}{0.9\linewidth}
\begin{grammar}
<type> ::= <id> 
\alt <stringLiteral> 
\alt <decimalLiteral>
\alt <realLiteral>

<decimalLiteral> ::= ( '0'-'9' )$^+$

<id> ::= ( 'a'-'z' 'A'-'Z' ) ( '_' 'a'-'z' 'A'-'Z' '0'-'9' )$^*$

<stringLiteral>  ::= '\"{}' ('_' 'a'-'z' 'A'-'Z' '0'-'9')$^*$ '\"{}'

<realLiteral> ::= ( '0'-'9' )$^+$ '.' ( '0'-'9' )$^+$
\end{grammar}
 \end{boxedminipage}
 \end{center}
\end{footnotesize}

In this section, we explored the design of the DSL through the different important parts defined in the language. 
In the following section, we explore the fourth step of the methodology for developing a DSL, i.e. implementation.

\subsection{Implementation}
The implementation is the development process of the DSL. It is based on the grammar of the language, which has been defined in the previous section.
In our case, we chose to start with an interpreter implementation. The main reason for this choice is to allow our language to be as efficient as possible by keeping syntax and semantics as close as possible to expert knowledge.
It is also easier to build good error reporting. Indeed, not relying on an underlying implementation allows to think deeply about the design and syntax of the language. This allows to be really in the domain specific, as named in "DSL". Moreover, it allows to get rid of existing error reporting systems to implement a language-specific one, for example by creating a semantic analyser that checks the typing of variables.

\subsubsection{Language Implementation}

Our language has been developed in java using the parser generator ANTLR\footnote{ANother Tool for Language Recognition: \url{https://www.antlr.org/}}.
ANTLR is a tool widely used to produce parser that can be used to explore Abstract Syntax Tree (AST).
Once this grammar is analysed by ANTLR, we get a tree parser with several possibilities to implement tree parsing, the visitor and the listener mechanism. The next step is to provide it with the corresponding language input to make its interpretation.

\subsubsection{Language Interpretation}
For linguistic interpretation and tree analysis, we chose to use the Visitor Design Pattern. The visitor allows a more complete exploration of the AST in regards to the semantics. For example, it will omit certain nodes and children or loop a certain number of times. 

Our interpreter has been designed to parse the tree to build an object representation of the input language. It is then this representation of objects in memory that is executed to perform the FDIA simulation. All the rules presented in the design section can be found in the implementation, either as a class or as an attribute.
This has several advantages. The first is the ease with which memory objects can be manipulated through the code before execution rather than directly during interpretation at the DSL level. This allows better code scalability. The second is related to the underlying tools. This makes it possible to implement and build scenario designs in a graphical way.

\subsection{Data Management}

As it has already been discussed above, data are the central element of FDIA and therefore of our approach. Also, IoT has hardly been standardised and in any case has been rarely used by manufacturers. This lack of standardisation is directly reflected in the data, whether in the data format used or in the data structure. For our approach, we would like to be able to carry out FDIA simulations on as many IoT as possible. Some requirements for data management are therefore to be defined: we should process data from as many IoT as possible, regardless of the structure and data types. The process should not lose any information. The input data must be able to be re-injected into the SUT without loss of information or format. Without this condition it would be complicated to inject consistently the altered data into the SUT, firstly because of format compatibility and secondly because intrusion detection tool would check the integrity of input data.

Therefore, we have set up a data management system allowing the manipulation regardless of the original format coming from the SUT. This makes it possible to perform data alteration in our tool independently of the format. The most efficient way to accept the majority of formats is to define a data format at the input of our tool.
For this purpose we decided to transform all input formats into a JSON format without structure. This is done using a flattening algorithm. The data structure is kept in the key part of the JSON key/value pairs representation. This allows us to retrieve the original format from the tool output thanks to an un-flattening algorithm. 

Once the input file is in the right format, it has to be saved in a database so that the data can be processed efficiently. The choice was made to use a document-oriented database, which does not require a pre-established database structure as MongoDB. These characteristics are interesting in our case, because we can process data of heterogeneous composition and structure. In this database one mongoDB document represents one complete message of one IoT device. Therefore, when querying a particular message attributes, you can act on the whole message through its representative document. A structure is nevertheless present in each document in the database. The information contained in the messages is placed under a BSON (Binary JSON) object: "properties". This allows the use of serialisation-deserialisation principles at the code level, which greatly facilitates the manipulation of documents and objects from the database.

In the next section, we explore different scenarios in detail for experimentation purpose.

\section{Experimentation}

Through this section, we aim to illustrate the grammatical expressiveness of the approach. We propose two examples: the first one performs a simple FDIA and analyses the result directly on the data, the second one is a more complex FDIA, which is similar to an actual use case.

To develop and test this framework, we use a real IoT system in production in the streets of multiple cities around the world. This system is used to monitor environmental conditions within cities, such as temperature, humidity, noise, particle concentration and nitrogen dioxide concentration.

The data schema returned by the devices is given in Fig.~\ref{fig:dataschema}. Data are retrieved every 15 minutes and sent immediately when the network is available.

\begin{figure}[!ht]
\centering
\begin{lstlisting}[language=json,firstnumber=1]
{"data": {
      "meter_code": "10",
      "temperatureTC": 8.03,
      "HumidityRH": 94.77,
      "LAeq":"6500",
      "No2":"24",
      "noise":[0,2,23,26,.....,44,33,22],
      "particles":18939,
      "location":"47.213865,5.968195",
      "timestamp": 637458300
    }
}
\end{lstlisting}
\caption{Data gathered by a thing in JSON}
\label{fig:dataschema}
\end{figure}

In the following experimentation, we work on data file extracted from the IoT system. Those files represent one month's worth of records.

\subsection{Experimentation 1}

 In this experimentation, we realise a simple FDIA to simulate a sensor malfunction or failure. This type of attack creates a basic disruption which is easily detected by a human, leading to costly maintenance decisions for the system operator, in addition to breaking the confidence of system operators and users.
 
\begin{figure}[!ht]
\centering
\subfigure[Initial temperature]{\label{fig:tempmai2019}
\includegraphics[trim={0cm 0cm 6cm 1cm},clip=true,width=0.23\textwidth]{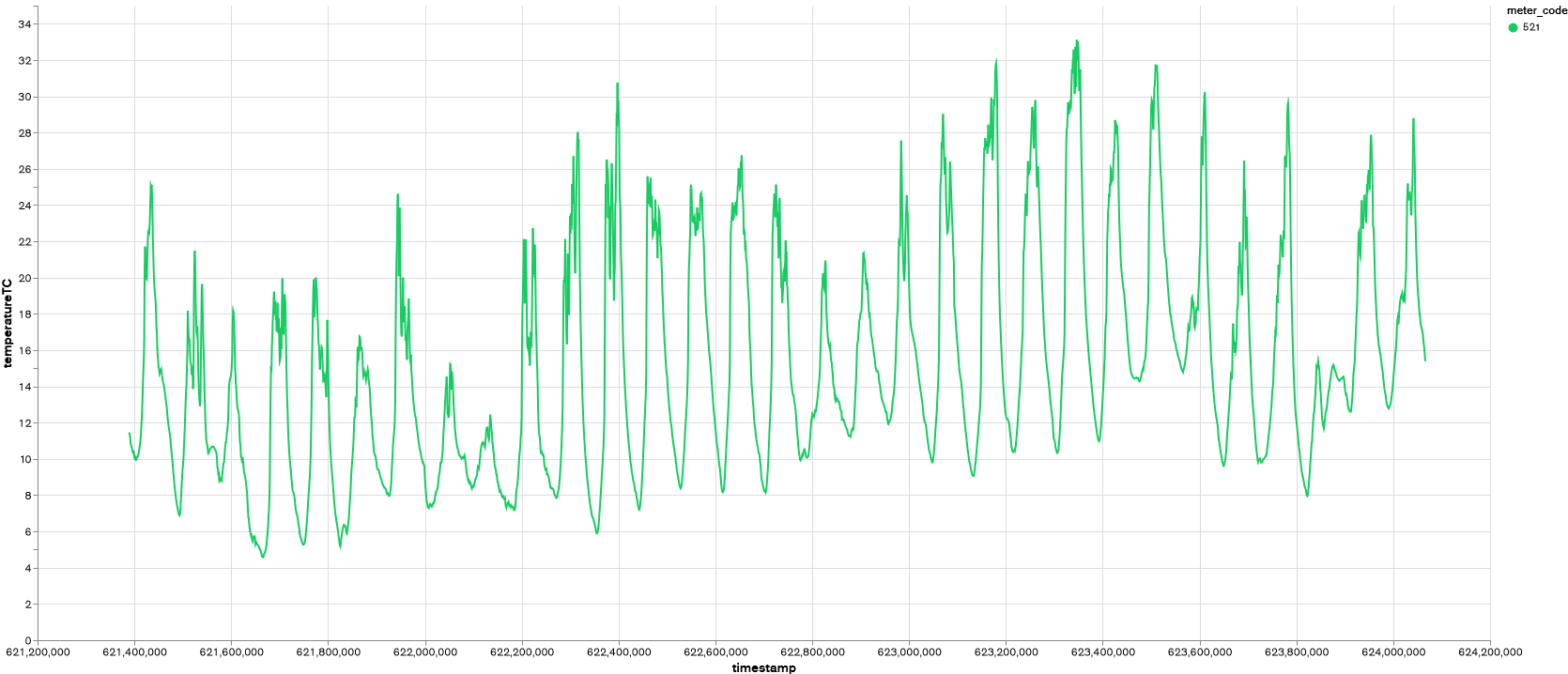}}
\subfigure[Altered data]{\label{fig:tempmai2019failsensor}
\includegraphics[trim={0cm 0cm 6cm 1cm},clip=true,width=0.23\textwidth]{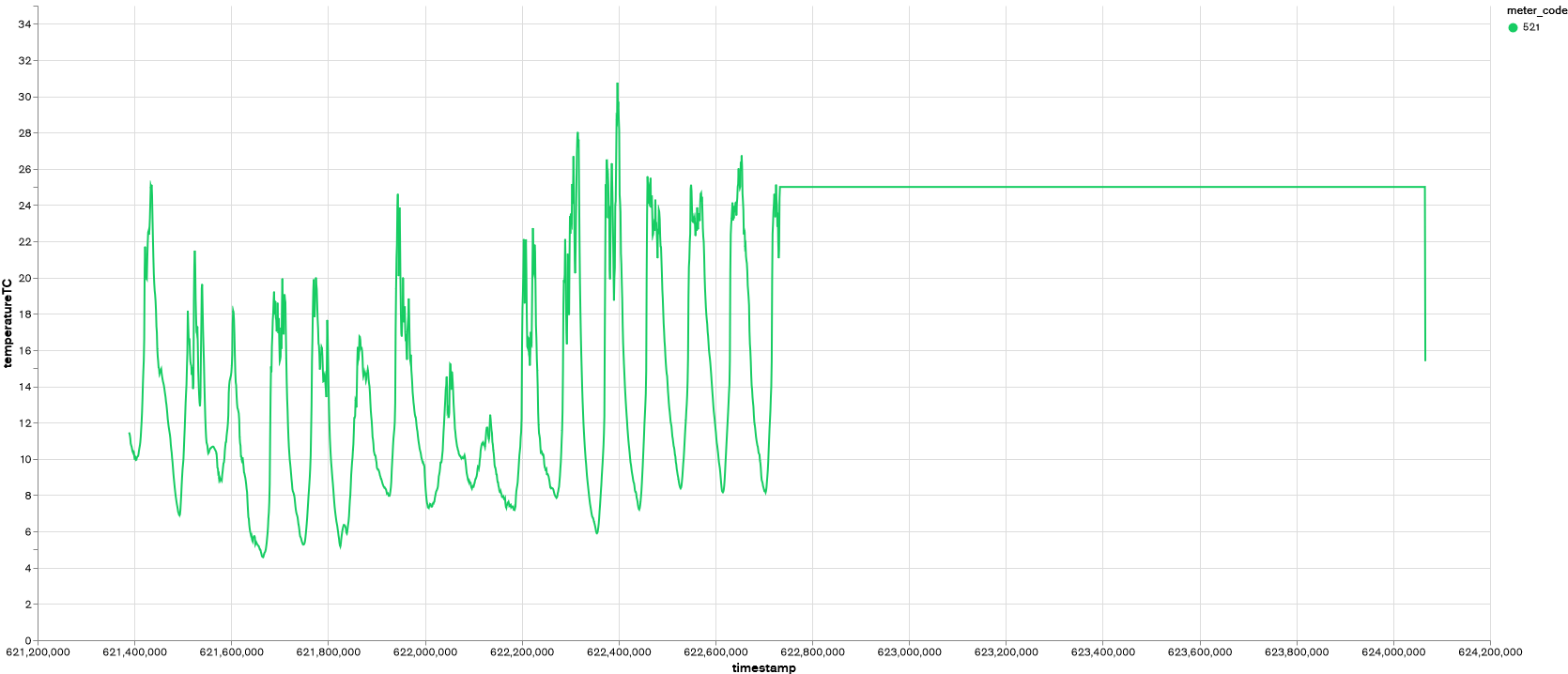}}
\caption{FDIA on Temperature in degrees Celsius in May 2019}
\end{figure}

 Fig.~\ref{fig:tempmai2019} shows the temperature data in Celsius in May 2019. Each spike on the line chart corresponds to one day. We apply the DSL scenario shown in Fig.~\ref{fig:FDIA1} to this dataset. The scenario describes an alteration where all the data in the time window defined are replaced by the value of 50 degrees Celsius. Fig.~\ref{fig:tempmai2019failsensor} shows the execution result. Visually any human can detect this attack, as well as any algorithm.
 
 With this experimentation we validate the simple selection and alteration of the grammar. We can produce the same kind of alteration on other data of the example. Such as the humidity to simulate the rain or on the number of particles and NO$_2$ in the air to simulate the absence of pollution or traffic jam.

\begin{figure}[!ht]
\centering
\begin{lstlisting}[language=fd2IoT,firstnumber=1]
scenario "failsensor"
ticker 2
alter things where meter_code="521" set temperatureTC=50 from 622732500 to 624066300;
\end{lstlisting}
\caption{FDIA for sensor disruption}
\label{fig:FDIA1}
\end{figure}

\subsection{Experimentation 2}

In this experimentation, we realise a FDIA with more complex selection and alteration function. The objective of this attack is to select multiple things through their geographical location, and to simulate a gradual increase in pm10 (particles between 2.5 and 10 micrometres) particle levels in a part of the city. For a more realistic attack, we need to select several sensors by their geolocation and apply to them the attenuation of the attack force according to their distance to the attack.
This type of attack can occur for several reasons, for example, to trigger alternating car traffic during pollution peaks or to lower real estate value in a neighbourhood.

\begin{figure}[!ht]
\centering
\includegraphics[trim={0.2cm 0cm 4cm 2cm},clip=true,width=0.45\textwidth]{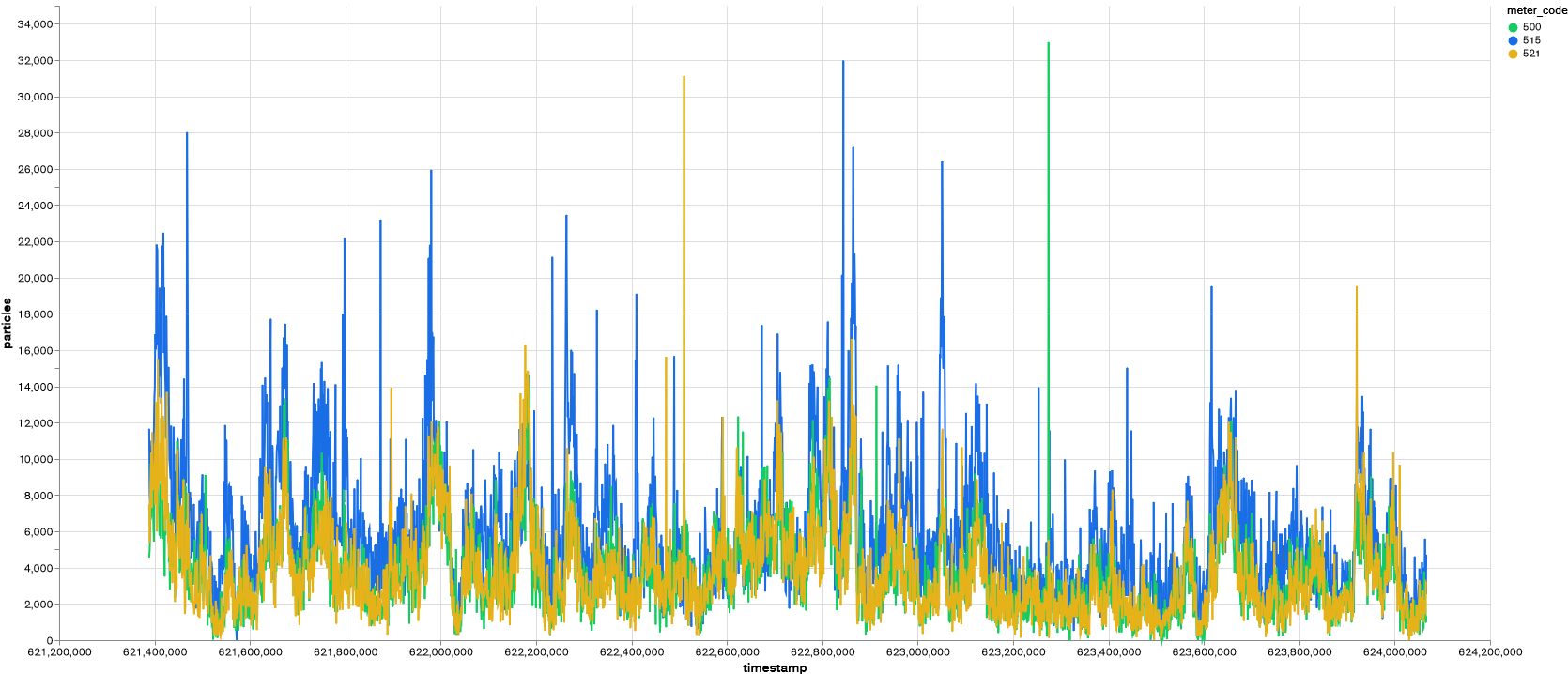}
\caption{Particles in may 2019 before an FDI attack}
\label{fig:particlebefore}
\end{figure}

For this experiment, we use data from the pm10 particles sensor gathered during the month of May 2019 by three things identified by their meter_code identifier: 500 (green), 515 (blue) and 521 (yellow). Chart of the data shown in Fig.~\ref{fig:particlebefore}.

\begin{figure}[!ht]
\centering
\includegraphics[trim={1cm 0.2cm 0.4cm 1cm},clip=true,width=0.4\textwidth]{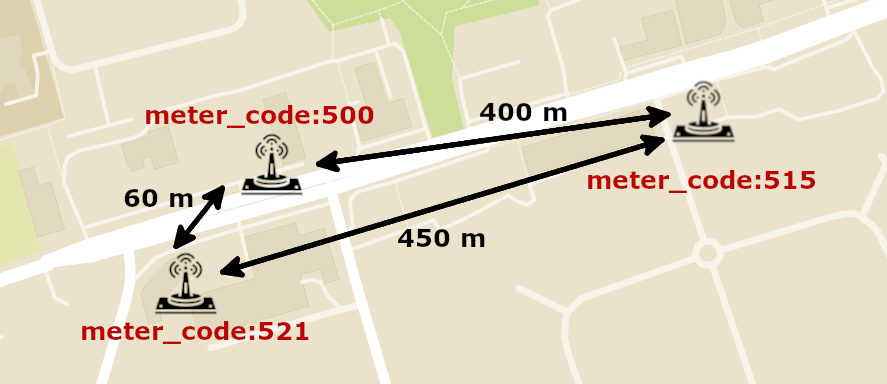}
\caption{Map of the sensor situation}
\label{fig:map}
\end{figure}

The scenario described in Fig.~\ref{fig:FDIA2} aims to select the three things through their geographical location, we choose to use the GPS coordinate of the thing 500 with a radius distance of 500 metres for the circle selection. So with this selection criterion we pick all the things present as shown on the map in Fig.~\ref{fig:map}, and we apply an increase to the selection that begins at 1 and increases over time with a step of 10. An attenuation of 10 by meter is applied to this alteration according to the distance from the application point of the scenario.

\begin{figure}[!ht]
\centering
\includegraphics[trim={0.2cm 0cm 4cm 2cm},clip=true,width=0.45\textwidth]{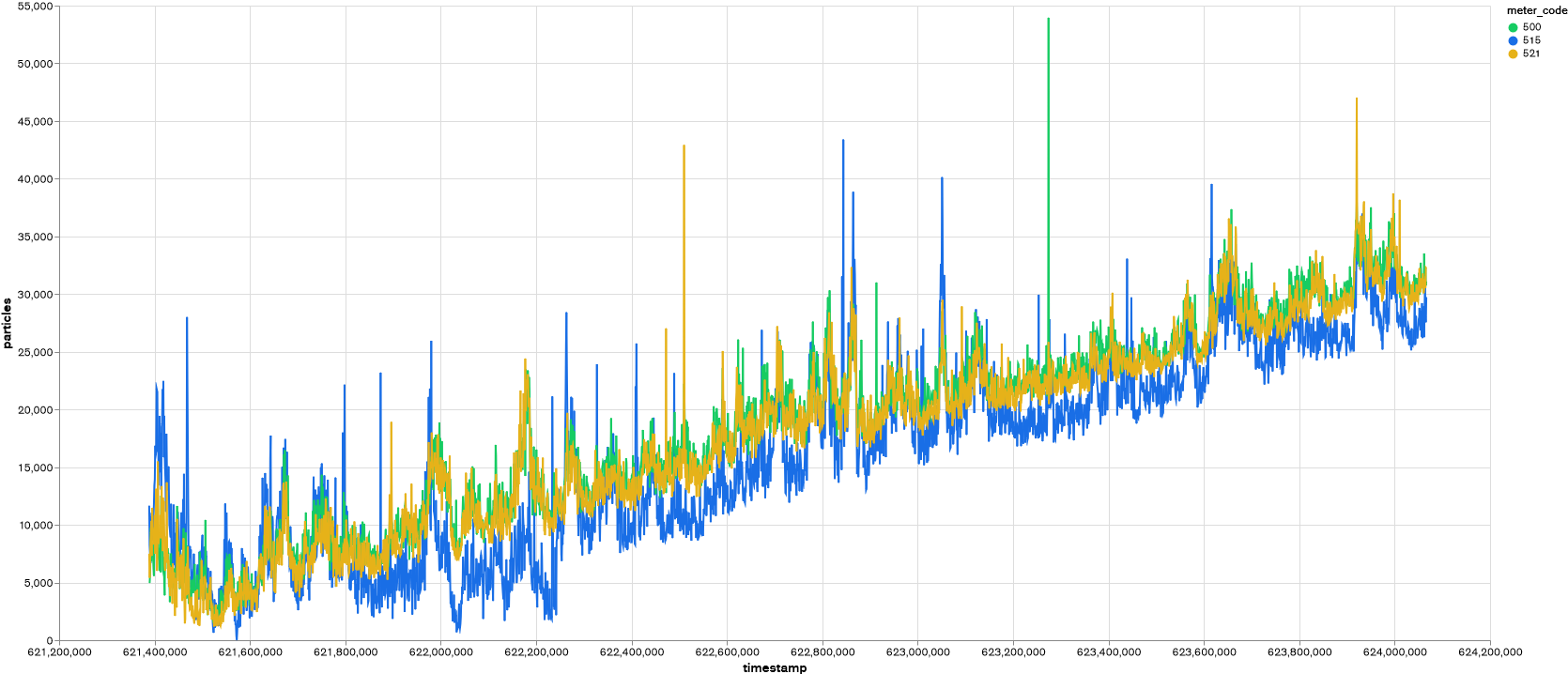}
\caption{Particles in may 2019 after FDI attack}
\label{fig:particleafter}
\end{figure}

Fig.~\ref{fig:particleafter} shows the attack result. We can see the effects on the three curves by the progressive increase of their values and that the farther the object is from the thing 500 (centre of the attack) the lower the value of its curve.
In the initial data, sensor 515 was predominant in terms of particle number, and sensors 500 and 521 were following almost the same trend. After the attack, the sensor closest to the attack (500) is predominant over the two others. Sensor 515 which is the furthest away detects fewer particles, this is visible in the figure since its curve is below the others.

\begin{figure}[!ht]
\centering
\begin{lstlisting}[language=fd2IoT,firstnumber=1]
scenario "IncrementationAndAttenuation"
ticker 2
geolocation (47.213865,5.968195)
alter things where location isInsideCircle(47.213865,5.968195,500) set particles+=(0.0->99999.0,10.0) with attenuation of 10.0 from 0 to 999999999;
\end{lstlisting}
\caption{FDIA for particle increase with distance attenuation}
\label{fig:FDIA2}
\end{figure}

The expressiveness of the approach makes it possible to define this type of complex attack on several sensors. For example, we can also simulate the propagation of sound in the city or evaluation of the heat and humidity level in the city to trigger a heatwave plan.


This section has demonstrated the usefulness and the performance of the tools provided. For experimentation 2, we have worked on 3 different connected objects which have produced 8931 records over the period of one month. The DSL scenario performed on this data was executed in under 2 seconds. 

\section{Conclusion}

In this work, we have addressed the FDIA challenge by proposing an approach and a language to describe an FDIA attack and then execute it on production data.
The prototype proposed in this article currently supports the processing of data from all types of IoT devices by help of JSON format flattening technique. It allows simple alterations such as assignments, but also complex alterations such as distance-dependent attenuation of alterations. Finally, it allows the export of these altered data in original format.
This approach can then be used for several purposes as to test the resilience of a system or to train machine learning tools to detect such attacks. 
As future work progresses, the DSL should be improved on its expressiveness, in particular, with more alteration and selection functions. Also, a link between the tool and SUT would allow performing tests directly from the tool. 


\end{document}